**RESEARCH**

# DHR+S: distributed hybrid rendering with realistic real-time shadows for interactive thin client metaverse and game applications

Yu Wei Tan[1] · Siang Ern Low[1] · Jonas Chow[1] · Javon Teo[1] · Anand Bhojan[1]

**Abstract**
Distributed hybrid rendering (DHR) is a real-time rendering approach that incorporates cloud-based ray tracing with locally rasterized graphics for interactive thin client metaverse and game applications. With cloud assistance, DHR can generate high-fidelity ray-traced graphics contents remotely and deliver them to thin clients with low graphics capability, including standalone extended reality devices and mobile phones, while maintaining interactive frame rates for users under adverse network conditions. DHR can already achieve the effect of ray-traced hard shadows that form with the occlusion of direct illumination. We enhance the realism of these shadows by softening their edges with the direction of rays traced and approximating the occlusion of indirect illumination by reconstructing ray-traced ambient occlusion with a modified version of spatiotemporal variance-guided filtering. Our technique uses only 20–30% of the bandwidth of remote rendering and is also tolerant of delays of up to 200 ms with only slight distortion to the shadows along object edges.

## 1 Introduction

Ray tracing is a realistic rendering technique that models the transport and approximates the properties of light in the real world. An alternative to ray tracing is rasterization which performs a poorer approximation as it does not simulate the bending behaviour of light as well as ray tracing. Nonetheless, this makes rasterization computationally lighter than ray tracing and, in general, the better choice for interactive applications with tight performance constraints like games.

Hybrid rendering [2] combines ray tracing and rasterization to achieve a performance-visual quality tradeoff in real-time rendering. Recent years have seen a speed-up in ray tracing hardware acceleration for personal computer (PC) GPUs like the NVIDIA GeForce RTX and AMD Radeon RX, which have made real-time ray tracing a reality for interactive applications on PCs with hybrid rendering. However, thin clients like standalone extended reality devices and mobile phones are still lagging in terms of graphics capability, and understandably so due to their form factor constraints such as heat generation and power consumption. Nonetheless, this makes ray tracing run too slowly for interactive frame rates. However, given the emerging demand for photorealistic and immersive metaverse and gaming experiences, it is worthwhile to try and bridge the graphics fidelity gap between such thin clients and PCs.

Instead of performing all the rendering in the clients, our previous work distributed hybrid rendering (DHR) [29] performs rasterization in the clients while leveraging remote PC servers for hardware-accelerated ray tracing, delivering the ray-traced output to clients over the network as a potential solution to achieve real-time hybrid rendering in standalone extended reality devices and mobile phones. Support for 5G technology offering high bandwidth and rapid data transmis-

✉ Yu Wei Tan
yuwei@u.nus.edu

Siang Ern Low
e0388996@u.nus.edu

Jonas Chow
e0544170@u.nus.edu

Javon Teo
e0725706@u.nus.edu

Anand Bhojan
banand@comp.nus.edu.sg

[1] School of Computing, National University of Singapore, Singapore, Singapore

sion speeds has arrived at such devices. Furthermore, the integration of the 5G multi-access edge computing (MEC) architecture is particularly noteworthy as it brings the remote server to the edge node (cloudlet or gamelet [3]) which is just 1 or 2 hops away from the client, allowing for ultra-low latency between the server and client. The high bandwidth and low latency capabilities of 5G MEC make it especially suitable for real-time hybrid rendering on thin clients which benefit greatly from cloud assistance when having to carry out resource-intensive computation. However, 5G MEC is currently only offered by a limited number of wireless service providers around the world. Nonetheless, many service providers are designing and conducting trials on MEC with 5G [38] so the infrastructure should pervade the market very soon.

5G uses radio waves for signal transmission. Unfortunately, radio propagation is susceptible to obstacles which are aplenty in the real world. Radio waves can reflect off or refract through surfaces, diffract around buildings as well as scatter through or get absorbed by atmospheric particles like dust. For indoor settings, the thickness and material of many types of walls also cause them to absorb radio waves, blocking the waves instead of allowing them to pass through. For a thin client performing DHR, an inconsistent network connection would result in unstable frame rates, leading to a choppy user experience which could not only negate the high fidelity of hybrid-rendered graphics but even worse, make certain interactive applications completely fail. For example, many first-person shooter games that require high speed and accuracy may become virtually unplayable. Hence, it is imperative to render up-to-date video frames albeit of lower graphics fidelity. DHR falls back on pure local rendering with rasterization only but also approximates the late or lost ray-traced data from the server for uninterrupted real-time hybrid rendering.

Diffuse reflection refers to the scattering of reflection rays off a surface and is commonly modelled as Lambertian reflectance where each scattered ray carries the same amount of light energy. This Lambertian shading model can be computed with rasterization but DHR augments its effect with ray-traced shadows. However, these shadows always have sharp edges while real-world shadows can have soft silhouettes (penumbra) given by the outward falloff of shadow intensity from the innermost darkest region (umbra) of shadows as they account for the light emitting surface area of light sources. Additionally, DHR casts rays towards light sources, taking into account the occlusion of direct illumination or emitted light. However, light is a wave in reality. Hence, the occlusion of indirect illumination or reflected, refracted and scattered light should also be considered. As such, the main contributions of this work are:

- Realistic real-time interactive shadow effects
- soft shadows for shadow penumbra
- ambient occlusion for indirect illumination
- Evaluations over a 5G edge computing testbed
- visual quality in terms of SSIM
- performance in terms of latency

## 2 Related work

### 2.1 Local, distributed and remote rendering

Distributed rendering (or collaborative/split rendering) is a polyseme: the division of rendering workload between computational resources depends on the context of use. For clarity, we refer to distributed rendering as the approach of having the client perform *some* but *not all* of the rendering. This is in contrast to local (client-side) and remote (server-side) rendering where the client performs *all* and *none* of the rendering respectively. Inspired by hybrid rendering, Tan et al. [29] provide a substantial review of real-time techniques in distributed and remote rendering, omitting local rendering to target thin clients for DHR. We add state-of-the-art works that have emerged since.

Several recent works on distributed rendering [18, 24, 33] consider multi-client setups where two or more clients display the same scene simultaneously, so they collaborate on the rendering to avoid repeating rendering computations. However, the performance of such approaches is heavily dependent on the number of available clients to collaborate with. For example, some clients may not be located near any other client for collaboration to be feasible while maintaining interactive frame rates. On the other hand, two close clients could still be accessing different parts or variations of the same scene, like when dynamic scene objects move differently. In such cases, distributed rendering might devolve into local rendering which would make it too slow for thin clients to display high-fidelity output in real-time.

In contrast, the division of rendering workload between multiple servers has also come up in recent years [9, 20, 22]. Despite the collaboration in rendering, such approaches cannot be classified under distributed rendering as the client does no rendering at all but just displays the rendered output generated by the servers. The performance of such approaches is hence fully dependent on the number of available servers to perform the rendering. Although this approach bestows application developers more control over the rendering pipeline with the rendering servers, there is still no failover to provide a temporally coherent user experience even in unstable network conditions. Additionally, edge sites tend to be small (with 1–3 servers) due to cost, maintenance and power constraints. Hence, approaches that only use a single server for rendering would be more practical considering real-life MEC systems. For example, two recent works choose to divide the rendering workload between the server and client in the 3D [21] or 2D [4] space, performing the rendering at different resolutions for objects or pixels respectively.

## 2.2 Soft shadows and ambient occlusion

For each pixel to be shaded, DHR casts a shadow ray from its world space position to the centre of every light source, forming hard shadows wherever direct lighting is occluded by some object. Nonetheless, the process of adding shadow penumbra and the occlusion of indirect illumination to DHR is more complex than just a tweak to simple ray casting. Hence, we provide a collection of dedicated state-of-the-art real-time rendering techniques for such advanced shadow effects.

With the use of artificial neural networks, neural rendering [30] is showing great promise in generating high-fidelity graphics with less manual effort and shorter render times as compared to traditional rendering, motivating recent works on soft shadows [7, 8, 25] and ambient occlusion [13, 17, 19, 37]. However, the execution of trained machine learning models on thin clients is generally too slow to meet the performance requirements of interactive applications. For instance, standalone virtual reality headsets would need to undergo fundamental changes to their underlying system-on-chip architecture to run such models efficiently [12].

Scene discretization techniques like signed distance fields [28] and spatial hashing [10] have also been employed to speed up the computation of soft shadows and ambient occlusion. Nonetheless, such techniques tend to run into issues with large scenes and thin objects when the resolution of the scene representation is too low while increasing its resolution would significantly increase its memory consumption or generation time.

Several advanced ambient occlusion techniques have also surfaced, albeit introducing additional complexity in implementation. Conical ray culling [36] optimizes the runtime computation of ambient occlusion involving dynamic objects by performing offline precomputation of occlusion between static objects and only tracing rays intersecting the bounding spheres of dynamic objects. On the other hand, stereo HBAO+ [26] reduces the stereo inconsistencies of ambient occlusion in virtual reality while SD-HBAO+ [32] takes into account indirect illumination occluded by hidden geometry.

## 3 Design

DHR achieves the effect of illumination occlusion from light sources that emit light from a single point. We boost the realistic quality of shadows by making them more scene-aware, taking into account the light emitting surface area of light sources as well as the potential occlusion of indirect illumination by nearby objects, providing easy user-controllable options to trade off between performance interactivity and visual quality.

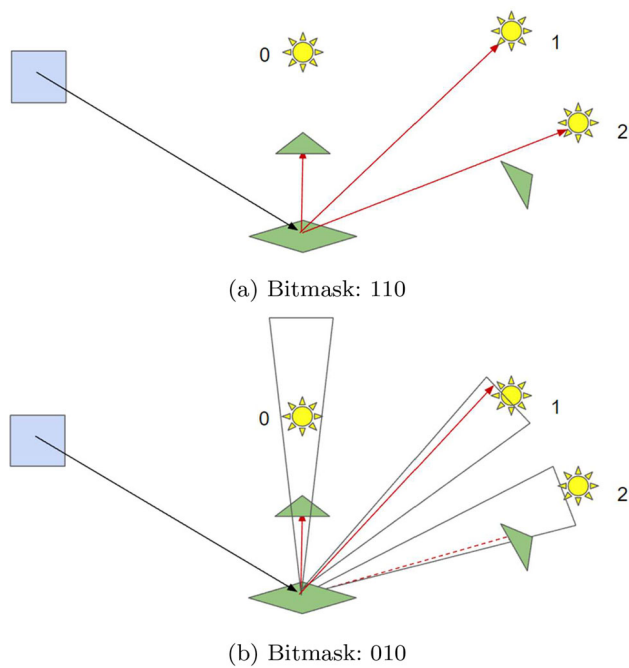

(a) Bitmask: 110

(b) Bitmask: 010

**Fig. 1** Shadow ray generation

### 3.1 Soft shadows

Instead of casting the shadow ray towards the centre of each light source like in Fig. 1a, we adopt the approach in Barré-Brisebois et al. [1] to apply a random but bounded offset to the direction of this ray while keeping its magnitude unchanged. Essentially, we are introducing directional variance to the ray within the solid angle of the light source forming a circular cone distribution that covers the entire light emitting surface area of the light source, as shown in Fig. 1b. Hence, for pixels near the edges of occluding objects, some shadow rays will hit the light source while others will hit nearer geometry, simulating the effect of partial occlusion and hence forming shadow penumbras corresponding to the amount of light emitting surface area.

The effect of this ray offset is seen from the resulting visibility bitmasks for the pixel corresponding to the scene point to be shaded. Light 0 is occluded in both cases as the triangle completely slices through the cone, while Light 1 is unoccluded as the ray does not hit any intermediate object while travelling the entire distance from the scene point to the light. Initially, Light 2 is not occluded from the pixel as seen in Fig. 1a. However, with the new approach, the randomly selected ray direction uncovers the presence of a nearby object as an occluder, resulting in a different visibility bitmask which is nonetheless of the same size as the original.

## 3.2 Ambient occlusion

From a scene point, it is easy to compute the direction of any direct illumination ray cast from a specific light source. However, indirect illumination rays cannot be computed just from the scene point itself as we do not know how light interacts with the scene before reaching the point to be shaded. Nonetheless, it is possible for us to simulate the interactions of reflection and refraction by *tracing* the path of *rays* from the camera to the scene point, then all the way to the illuminating light source, making use of laws of optics to calculate the necessary angles of reflection and refraction along the way.

However, this approach of tracing indirect illumination rays recursively throughout the scene [34] is computationally expensive. Hence, ambient occlusion [39] came about as an alternative model for simulating indirect illumination by approximating the effect of recursive ray tracing without its expensive computations. Ray-traced ambient occlusion (RTAO) [15] casting multiple "rays" of bounded magnitude within a hemisphere (as compared to a single ray within a cone for soft shadows earlier). The idea is to perform a local inspection of nearby objects which can block reflected, refracted or scattered light from reaching the point to be shaded. Scene objects are grouped hierarchically into bounding volumes of simple shapes like spheres, making it efficient to test for ray intersection. If a ray does not intersect a bounding volume, we know that it does not intersect any bounding volumes or objects within it. Given that the lower the magnitude of a ray, the lower the probability that it hits a bounding volume, bounding the magnitude of rays reduces the depth of tree traversal of the bounding volume hierarchy.

Ambient occlusion is usually defined as the cosine-weighted fraction of the hemisphere that occludes the surrounding ambient light [14]. Even though Laine and Karras [14] notes that this simple definition is usually unsatisfactory, we have found that this definition provides us with adequate results and minimises the data size sent over the network. Therefore, the ambient occlusion model used in our pipeline can be represented by Eq. (1).

$$O(\mathbf{p}, \mathbf{n}) = \frac{1}{\pi} \int_{\Omega_r} \omega \cdot n \, d\omega \quad (1)$$

As the integral is analytically complex to compute, we can approximate it by sampling the hemisphere with a number of rays. The rays are shot using a cosine-weighted distribution which allows us to remove the dot product in the integral,

making the equation:

$$O(\mathbf{p}, \mathbf{n}) \approx \frac{1}{N} \sum_{i=1}^{N} \rho(\mathbf{p}, \omega_i) \quad (2)$$

$O(\mathbf{p}, \mathbf{n})$ is the amount of occlusion around the surface point $\mathbf{p}$ with a surface normal $\mathbf{n}$. $\Omega_r$ refers to the hemisphere oriented towards the normal $\mathbf{n}$ with radius $r$. With this formulation, we can tweak the amount of occlusion received by changing the value of $r$. Any ray that intersects with a primitive at a distance less than $r$ will be treated as occluded, while any ray intersection with distance larger than $r$ is ignored. Here, $\rho(\mathbf{p}, \omega_i)$ equals 1 if the ray $\omega_i$ is unoccluded (i.e. ray intersection occurs beyond the hemisphere's radius) and 0 otherwise. This gives us an occlusion factor that ranges from 0 to 1. We therefore have the following formula:

$$O(\mathbf{p}, \mathbf{n}) = \frac{N_{\text{unoccluded}}}{N}, \quad O(\mathbf{p}, \mathbf{n}) \in [0, 1] \quad (3)$$

The retrieved ambient occlusion is then applied as an attenuation factor onto the final pixel colour. We first do ray-tracing on the server to calculate $N_{\text{unoccluded}}$ for each pixel using the aforementioned method. We then send this data over to the client which calculates the final occlusion factor and combines it with the final pixel colour. Due to the real-time requirements, we can only shoot a low number of ambient occlusion rays to keep the application interactive. Therefore, the data size used for each pixel is only 1 byte as the total rays shot per pixel for ambient occlusion should not exceed 255, allowing us to store $N_{\text{unoccluded}}$ in an 8-bit integer.

The user can vary both the radius of the sample hemisphere and the number of rays shot per point to calibrate the ambient occlusion strength based on preference and balance the tradeoff between accuracy and performance. We can see from Fig. 2 that increasing the number of rays will increase the accuracy of the ambient occlusion with diminishing returns as number of rays increased. In particular, for this scene, SUN TEMPLE, there was no noticeable difference between using 32 and 64 rays per point. Similarly, varying the radius of the hemisphere affects the accuracy of the ambient occlusion as shown in Fig. 3. The larger the radius of the hemisphere, the more likely the sample rays will hit valid geometry and be occluded, resulting in a stronger ambient occlusion effect. Using these two parameters, we can produce distinctly different ambient occlusion effects while keeping the size of data sent the same.

## 3.3 Modified SVGF

Diffuse and glossy lighting effects are achieved via distributed ray tracing [6]. Such fuzzy-looking phenomena are problems of numerical integration over space and time. They can be solved using Monte Carlo integration through the stochastic sampling of scene points via multiple rays. A sufficient number of rays needs to be traced for a good approximation of the integral computed, but this is computationally expensive. For ambient occlusion, since the ray directions were shot at random, the resulting ray counts were noisy which led to poor compression rates. The larger compressed frames were more easily lost, affecting the frame rate (fps) of the application. Hence, we cast an insufficient number of ambient occlusion rays but apply a modified version of spatiotemporal variance-guided filtering (SVGF) [23] on the ray-traced ambient occlusion data at the server to denoise the output.

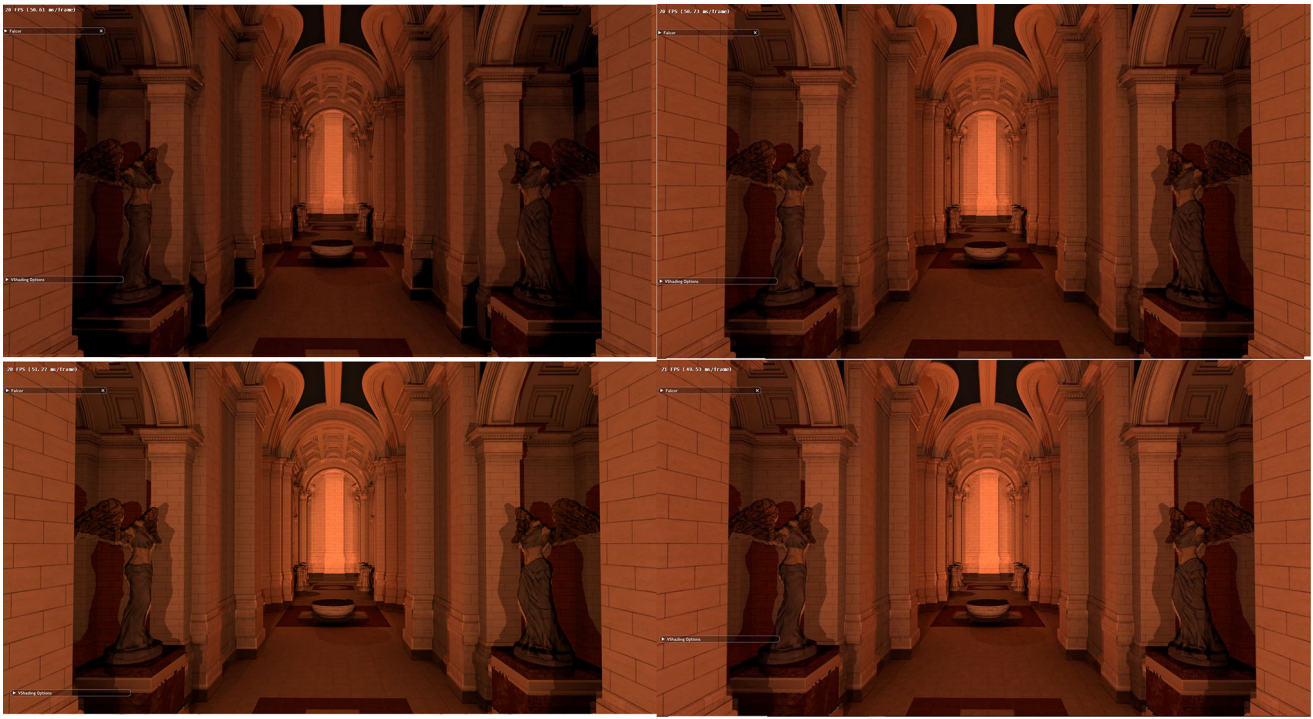

**Fig. 2** Different no. of ambient occlusion rays: 8 (top left), 16 (top right), 32 (bottom left), 64 (bottom right)

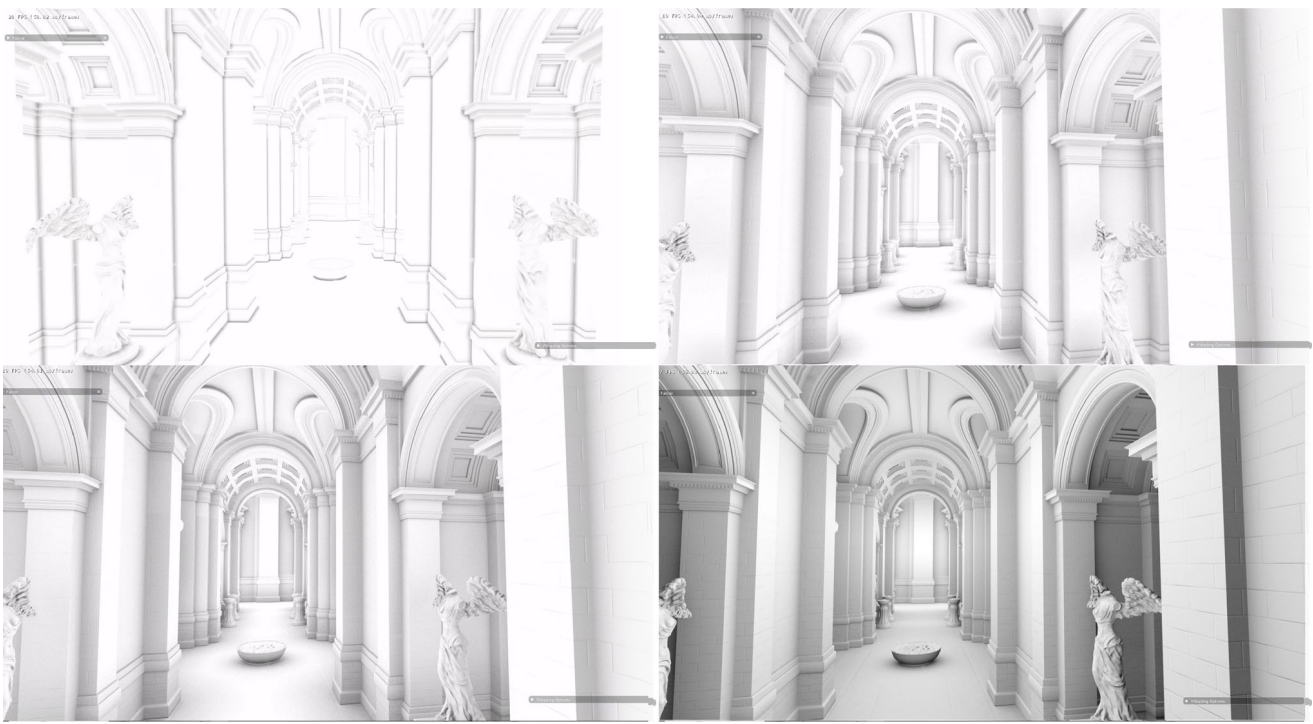

**Fig. 3** Different ambient occlusion hemisphere radii: 0.1 (top left), 1 (top right), 2 (bottom left), 5 (bottom right)

The original implementation of SVGF was used for global illumination using 1 sample-per-pixel (spp) rays for both direct and indirect illumination methods. The filter computes the light received at each pixel as a weighted average of its neighbours both temporally (using previous frames) and spatially (within the same frame):

- Temporally, the filter accumulates samples via an exponential moving average, integrating colour information from previous frames into the current frame.
- Variances in luminance are used to detect noise, allowing us to avoid filtering areas with low variance and prioritise areas with more noise.
- Motion vectors are stored within our current GBuffer, which allows us to backproject the current pixel's world position onto the previous frame.
- The filter samples the previous pixels to perform filtering. All invalid samples, i.e. samples that are inconsistent in terms of projected depths, normals and mesh IDs, are discarded.
- In cases where the previous frame history is not available (e.g. when the application starts or no valid samples are available), the filter instead performs spatial filtering.
- Spatially, the filtering is performed the same way. A larger kernel is used to sample the neighbouring pixels within the same frame to calculate the variance needed to perform filtering.
- We continue spatial filtering until we have accumulated enough temporal information to return to doing temporal filtering.
- The variance calculated then from the above steps drive the weights in the kernel for the iteration step.
- The current pixel has the kernel applied to it for multiple iterations, with each previous iteration refining the current iteration.

To modify the SVGF for filtering ambient occlusion, the variance computation were changed to no longer consider luminance as a factor. Instead, it uses differences in depth and surface normals to calculate the variance at each pixel. The results of the filtering are shown in Fig. 4.

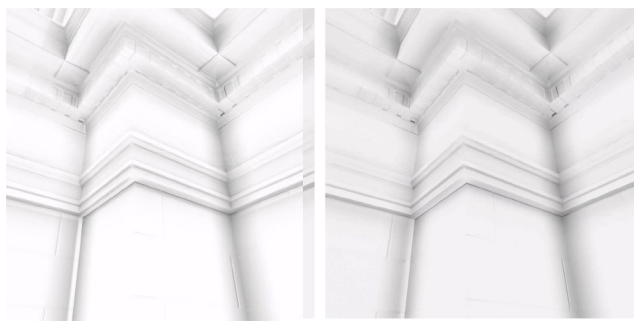

**Fig. 4** Ambient occlusion without (left) and with (right) SVGF

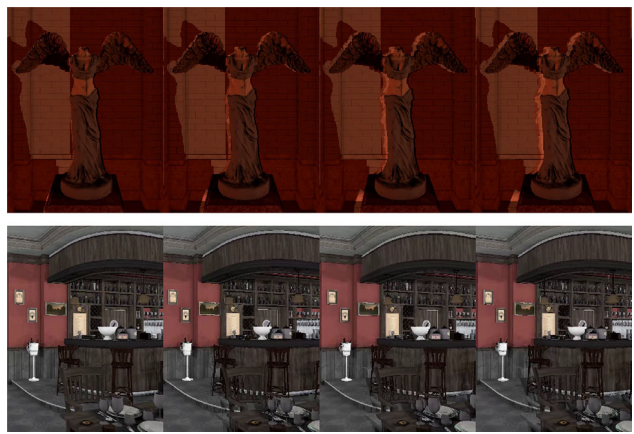

**Fig. 5** Lag (ms) from left to right: 0, 50, 100, 200

In the future, it is possible for us to switch to a more lightweight denoiser for ambient occlusion due to many of the pixels having a high number of temporal samples. This would entail doing just one iteration of filter and also using a smaller kernel size.

## 4 Evaluation

### 4.1 Latency simulations

To evaluate robustness of our hybrid rendering pipeline in the face of varying network conditions, we simulated a range of different network latencies by only sending the camera data to the server after a specified delay, from Fig. 5, we see that the application is tolerant of delays of up to 200 ms, with only slight distortion to the shadows in the frames. On the other hand, Fig. 6 shows the respective frames without prediction in place. The shadows are much more distorted and offset compared to the version with prediction.

Furthermore, due to the hybrid rendering nature of the pipeline, each of the lighting passes are independent of each other and could in theory be sent separately. This would allow us to use the most recently received frame either directly or to predict the current frame for each pass. On the other hand, for remote rendering, it is more difficult for the client to perform prediction since it only receives colour data of the final frame and not the data that makes up each of the lighting effects.

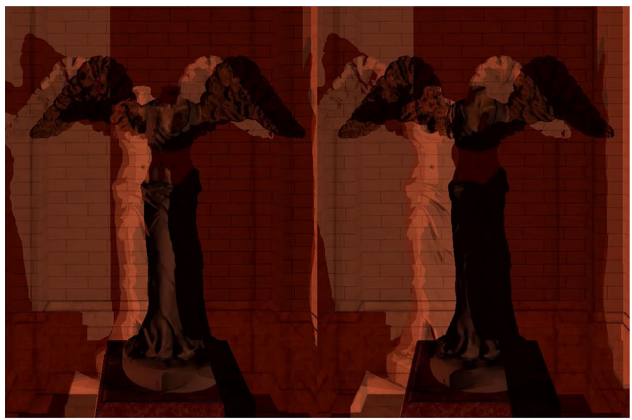

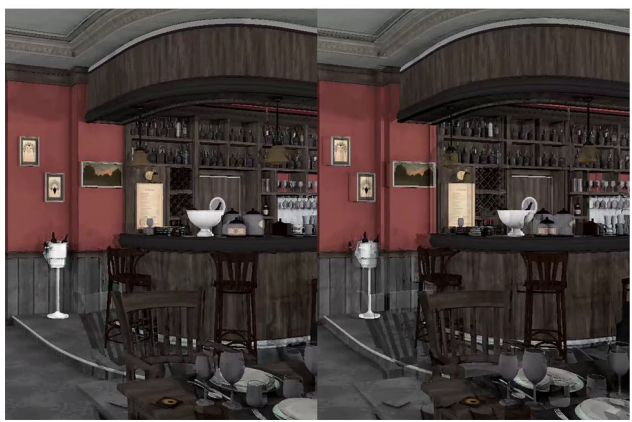

**Fig. 6** Without prediction: 100 ms lag (left) and 200 ms lag (right)

**Table 1** Bandwidths

| Bandwidth (Mb/s) | **Hybrid** | **Remote** |
| --- | --- | --- |
| PINK ROOM | 172 | 579 |
| SUN TEMPLE | 129 | 619 |
| BISTRO | 166 | 596 |

The bandwidth used by the pipeline is given by the total size of data transmitted over a given amount of time. For this evaluation, Wireshark was used to monitor the packets travelling from the server computer to the client computer. The application was allowed to run briefly to reach a steady state before Wireshark was used to capture and analyse the outbound packets from the server for a duration of 1 min. The bandwidth used was retrieved from the summary page provided by Wireshark and is detailed in Table 1.

The scenes used were PINK ROOM, UE4 SUN TEMPLE AND AMAZON LUMBERYARD BISTRO (CC BY- NC- SA).

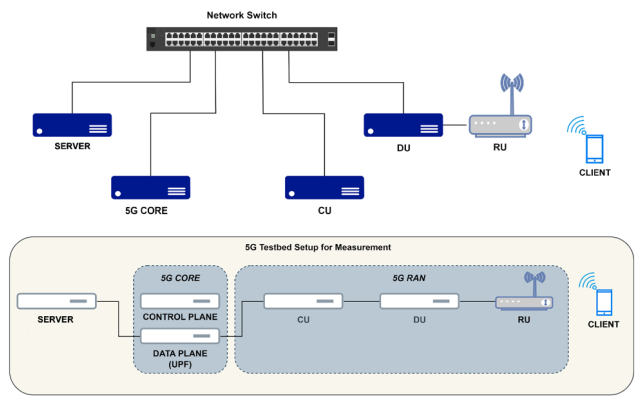

**Fig. 7** Physical (top) and logical (bottom) structure of 5G setup used

For all three scenes, we can see that the hybrid version uses much less bandwidth, around 20–30% of the fully remote version.

It should be noted that the scenes were tested with hard shadows and ambient occlusion, with the same radius and 32 rays shot for ambient occlusion for the three scenes. The bandwidth used for the hybrid version would increase if the ambient occlusion effect was strengthened. However, we can still expect that the hybrid version would use significantly less bandwidth than the remote version.

### 4.2 Real measurements on 5G testbed

We also tested the hybrid rendering version over a 5G network. Figure 7 shows the physical and logical structure of the 5G setup we used. The server is run in an edge node which directly connects to the base station, which the client can connect to. The data from the client flows to the distributed unit (DU) through the radio unit (RU) for radio processing, then passes through the central unit (CU) for higher-level processing, then reaches the 5G core (5GC) and finally the server.

We also measured the end-to-end latencies between the client and the server, as shown in Fig. 8. We have an average latency of 12 ms, and the 99th percentile is 48.3 ms. These latency values are at the lower end of the range of latencies considered in our simulations, and thus we can expect that the visual quality is very good. Assuming the client runs at 60 fps, a 12 ms delay would only add less than 1 frame difference between the server and the client, which should lead to a minimal difference from having no delay.

Figure 9 provides a visual comparison between the scene rendered with the hybrid rendering pipeline over a local network with no latency, and over the 5G testbed with an average of 12 ms of latency. As can be seen, the shadows between the two images are extremely similar, except for some fine details. We have attained an SSIM score of 0.876 when comparing between the two images, suggesting that they are extremely similar to each other. There was also no observable difference in the frame rate and responsiveness between the two setups.

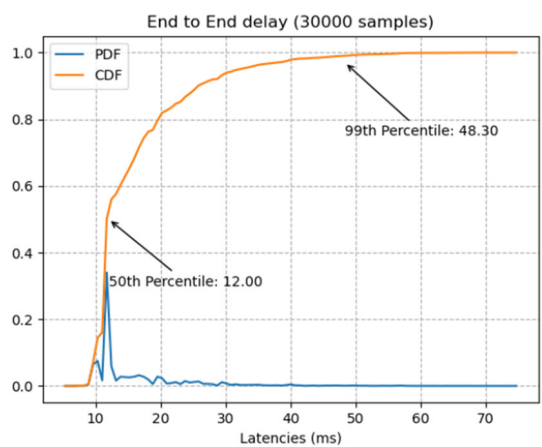

**Fig. 8** End-to-end latencies measured between the client and server

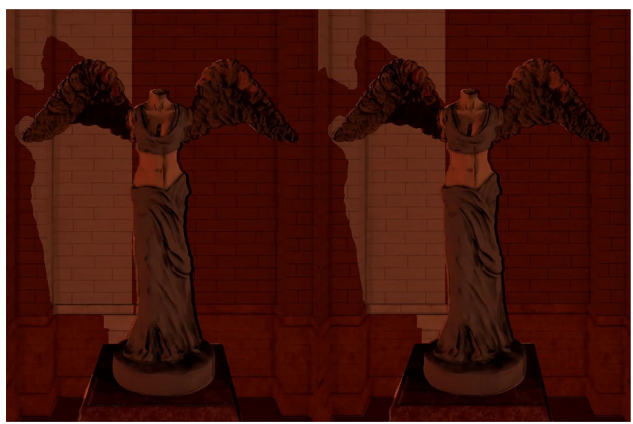

**Fig. 9** Comparison between local (left) and hybrid (right) rendering at 12 ms latency

### 4.3 Metrics

#### 4.3.1 Frame metrics

There are several points within the pipeline that can be optimised in order to improve performance. Since we are sending data from the server to the client, this data transfer is one of the main bottlenecks when it comes to rendering an interactive application. Therefore, we can measure both the size of the data sent per frame and also the speed at which this data is sent to the client to evaluate the performance of our hybrid rendering pipeline.

We look at the following three metrics: frames per second (fps), frame time and packet size. Frame time refers to the time taken to generate a single frame. Packet size here refers to the total size of the packets that make up a single compressed frame that is sent over the network.

**Table 2** Frame metrics

|  | FPS | Frame time | Packet size |
|---|---|---|---|
| No passes | 43 | 23 | 300,000 |
| Visibility pass (hard) | 44 | 22 | 230,000 |
| Visibility pass (soft) | 37 | 27 | 1,000,000 |
| Ambient occlusion | 37 | 26 | 1,600,000 |

The results were obtained by running the program and letting the scene come to a resting state, where there is little variance in the aforementioned metrics. The program was run using both hybrid and fully remote methods. Full remote rendering refers to having the server fully render the game frame and sending over the colour data. In such a setup, the client only needs to receive and display the frame and does not need any further processing. Soft shadows were implemented using a cone distribution described in Barré-Brisebois et al. [1], without filtering implemented.

Table 2 shows the results of the different metrics against different passes. Data shown is for the hybrid rendering version. It should be noted that each row of the table applies the pass to the row before it. For example, the row with Visibility pass (hard shadows) applies that pass on the previous, which has no passes, while the final Ambient Occlusion row has both soft shadows and ambient occlusion applied.

The remote rendering solution required almost a network with virtually no packet loss to be able to render the scene interactively, which is not possible in most networks in the real world. Hence, only the packet size of the remote rendering could be captured. Since it was the entire colour buffer of the final frame, it was independent of the number and type of passes, and the number remained at a steady 6,300,000 for all the passes.

Therefore, we can see that our hybrid rendering version leads to a significant drop in the size of data sent, with data sent per frame being around 25% that of the remote rendering version. We can see that applying both ambient occlusion and hard shadows does not result in a significant performance drop. The largest drop in performance stems from the soft shadows, which is expected as it is unfiltered and noisy at this stage. The final fps of all the passes is above 30 fps which could then be reprojected to 90 fps to fulfil the requirements of VR. The latency of the passes is also below the latency threshold (40 ms) for interactive gaming. This is promising as these implementations have areas that could be optimised further in the future, increasing the performance.

One area to improve these implementations are to decrease the packet size. The large increase in packet size for soft shadows is due to the noise generated by shooting the random shadow rays. Therefore, we can reduce this noise by applying a filtering the visibility bitmap, either via SVGF or another filter.

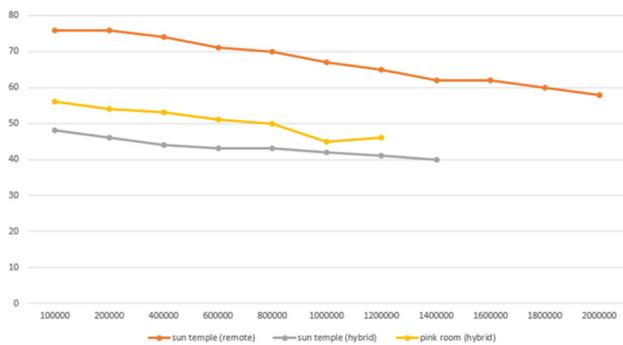

**Fig. 10** Graph of frame size versus fps

### 4.3.2 Frame size versus FPS

The term frame size in this section will refer to the compressed size of the frame on the server side. These compressed frames are split into equally sized packets which are then sent over the network as UDP packets.

From Fig. 10, the fps decreases as the compressed size of the frame sent over the network increases with a linear relationship. Furthermore, the smaller the frame size, the less likely it is for the frame to be dropped as less packets would have to be sent over the network. The relationship between frame size and fps on the client was evaluated by limiting the size of the frame sent over the network. As the frame sizes of the hybrid versions are much smaller, the data points obtained end earlier than the remote rendering version.

We can see that the remote rendering actually results in a higher fps that the hybrid rendering version. However, in reality, remote rendering is still not as feasible as hybrid rendering due to the large frame size generated.

As mentioned in the section on frame metrics, the remote rendering method used on the SUN TEMPLE scene has a frame size of around 6,300,000. By extrapolating the data above, we can estimate that the final fps for the remote rendering version to be around 17 fps, which is not feasible in a real-time rendering application. Furthermore, this also does not account for the fact that most of these frames would also be lost due to the large frame size, leading to the resulting scene to freeze after a while as there are no new frames received to redraw the scene.

One reason why the hybrid version has a lower fps compared to the remote rendering version is because the client has to do multiple copies of the data into the texture. In our current setup, the client has to copy the visibility and ambient occlusion data into their respective textures before drawing the final scene. These copies incur additional performance costs. However, the resulting pipeline is more resistant to adverse network conditions as shown in Fig. 9.

## 5 Future work

In addition to ray-traced shadows from direct illumination, we consider the ambient occlusion effect given by indirect illumination. We are also looking to add other lighting and camera effects such as reflections and depth of field to the pipeline. However, the increase in the number of data buffers for these additions introduces more complexity to the network components of the pipeline. In particular, more ray-traced data leads to an increased likelihood of lost or corrupted packets and hence the probability of frames being dropped, affecting the interactivity of the application. Hence, we can consider replacing the current LZ4 compression algorithm with a more efficient spatiotemporal one like H.264 encoding, and SVGF with more effective denoising algorithms.

Although machine learning is currently not that feasible on thin clients like standalone extended reality devices and mobile phones, we anticipate more hardware and operating system advancements down the line to run trained models on thin clients. Hence, we are still exploring its possibilities in denoising [31] and super-resolution [35] to reduce the number of rays needed to be traced as well as frame interpolation [11] to reconstruct frames with lost or corrupted packets to maintain the application's level of interactivity.

For soft shadows, Lauterbach and Manocha [16] first performs rasterization-based shadow mapping to identify pixels that are potentially inaccurate due to aliasing artifacts, then splats these pixels based on the shape of the light and their distance from it, forming a ray mask. The masked pixels are ray-traced as they are deemed to potentially be in some shadow penumbra, while the unmasked pixels marked in the shadow map are deemed to be in the umbra region. We are currently working on incorporating this a hybrid rendering effect that applies selective rendering [5] to avoid tracing shadow rays for every pixel.

Sun et al. [27] is a recent paper on generating soft shadows specifically for mobile devices, taking into account their limited computational capability and memory. We can also adapt some ideas to improve our shadow prediction.

## 6 Conclusion

We elevate the realistic quality of shadows in the DHR pipeline by implementing ray-traced soft shadowing and filtered ambient occlusion for indirect illumination, achieving higher performance stability and bandwidth savings as compared to remote rendering. Our work takes another step towards immersive real-time ray-traced graphics in interactive applications and games running on standalone extended reality devices and mobile phones.

**Acknowledgements** This work is supported by the Singapore Ministry of Education Academic Research grant T1 251RES2205, "Real-time Distributed Hybrid Rendering with 5G Edge Computing for Realistic Graphics in Mobile Games and Metaverse Applications".

**Author Contributions** Yu Wei Tan and Siang Ern Low wrote the main manuscript text while Jonas Chow and Jovan Teo did the performance evaluations. All authors reviewed the manuscript.

**Data availability** No datasets were generated or analysed during the current study.

**Declarations**

**Conflict of interest** The authors declare no competing interests.